\documentclass[twocolumn]{revtex4}

\usepackage{amssymb}
\usepackage{amsfonts}
\usepackage{amsmath}

\begin{document}

{\bf Comment on ``Why is the DNA denaturation transition first order?''}

Recently, Kafri, Mukamel and Peliti (KMP) \cite{KMP} extended the
classical Poland-Scheraga (PS) model for the denaturation transition of DNA
\cite{PS70}. According to PS, the energetic binding of bases in the double
helix competes with the entropic contribution of denatured loops, 
implying that the nature of the transition depends on the
exponent $c$ in the statistical weight $\Omega(2 k)\sim s^{k}
k^{-c}$ for a closed loop of length $2k$: for $1<c\le 2$ it is
of second order, while for $c>2$ it is of first order \cite{PS70,KMP}.
Fisher, taking the effects of self-avoidance within a
denatured loop into account, found $c=d\nu\approx 1.766$ in $d=3$,
i.e., a second order transition \cite{Fis66}.
In Ref.\ \cite{KMP} it was obtained that the exponent $c$ is modified
if additional effects of self-avoidance between a denatured loop and
the vicinal double helices are included. For a single loop within two 
strands of double helix, it was found that \cite{KMP}
\begin{equation}
\label{sex}
c=d\nu-2\sigma_3\approx 2.115 \, \, , \quad d = 3 \, \, ,
\end{equation}
i.e., a first order transition, where $\sigma_3$ is a
topological exponent related to a 3-vertex of a polymer network 
\cite{Dup86}.

This conclusion is valid in the asymptotic scaling limit for long flexible,
self-avoiding chains, i.e., each of the three segments going out from a
vertex must be much longer than the persistence length $\ell_p$ of this
segment (even though individual values of the $\ell_p$ may be different).
If this condition is fulfilled, the analysis for the PS-inspired model
in Ref.\ \cite{KMP} is consistent \cite{sim,M}. 
However, we point out in this Comment that it
does not apply to the chains typically used in experiments \cite{WB85}: the
DNA double helix being quite rigid, we expect the transition in such systems
to be of second order.

The typical length of DNA used in experiments varies from about 100 to 5000
base pairs (bp), the latter corresponding to a whole viral DNA \cite{WB85}.
In such DNA, a chain ``monomer'' $m$, which corresponds to a bead in a freely
jointed chain, represents one persistence length $\ell_p$. For the single
strand in a denatured loop, typically $\ell_p({\text{L}})\sim 40${\AA}
(roughly 8 bases), whereas for the double helix $\ell_p({\text{H}})\sim
500${\AA} (100\,bp) \cite{MS95}. Even if one assumes
that a segment of 10 monomers $m({\text{H}})$ of the double helix is already
long enough to be sufficiently close to the asymptotic scaling limit 
for long chains (which is hopelessly optimistic \cite{SFB92},
and also much less than taken in the simulations \cite{sim}), one can
{\em at best\/} place 5 loops even on the longest chains such that the
flexibility condition is not violated. However, with a maximum number of only
5 loops, the system is governed by finite size effects, and the analysis
in Ref. \cite{KMP} is no longer valid.

Conversely, only 80 bases (40\,bp) are needed to form 10
monomers $m({\text{L}})$ such that a denatured loop can be considered 
as sufficiently flexible. The longer chains in the experiments can 
thus exhibit a fairly large number of such loops, if the segments of 
the double-stranded helix between them are allowed to be of the order 
of $\ell_p({\text{H}})$ (see below).
It is therefore justified to neglect the entropy of the double-stranded 
helix \cite{PS70}. Following this picture, a vertex with three 
outgoing legs would not tie together three flexible chains, but rather 
two flexible chains (belonging to the loop) and one {\em rigid rod\/}
(the double helix). However, a rigid rod represents an 
irrelevant object for flexible, self-avoiding chains in the scaling 
limit for $d=3$ \cite{HED99}, which implies for the present case that 
$\sigma_3$ in Eq.\,(\ref{sex}) is replaced by zero. This, in fact, 
reproduces the original value $c\approx 1.766$ \cite{Fis66}, i.e.,
a second order transition.

Finally, we note that at the denaturation transition of real DNA
the average length $\xi$ of bound helical segments between loops is of the 
order of a few hundred bp, thus comparable with $\ell_p(\text{H})$ and much 
larger than $\ell_p(\text{L})$ (see, e.g., Eq.\,(9.129) in Ref.\,\cite{PS70}).
This large value of $\xi$ reflects the 
fact that the denaturation transition of real DNA is 
{\em highly cooperative\/};
in the PS model, this is generally modeled by the (non-universal) 
cooperativity 
parameter $\sigma_0 \ll 1$ \cite{PS70,WB85,Blake}
which enters the statistical weight of loops in the partition function,
leading to $\xi \sim 1/\sigma_0 \gg 1$ (see Eqs.\,(9.1), (9.129) and
(9.115a) in Ref.\,\cite{PS70}).

We thank M.~Kardar and H.~Scheraga for comments. This
work was supported by the DFG and by NSF grant DMR-01-18213.

\vspace{0.1cm}

\noindent
Andreas Hanke and Ralf Metzler\\
Physics Department, MIT, Cambridge, MA 02139\\

\end{document}